\documentclass[conference,letterpaper]{IEEEtran}
\IEEEoverridecommandlockouts

\usepackage[utf8]{inputenc} 
\usepackage[T1]{fontenc}
\usepackage{url}
\usepackage{ifthen}
\usepackage{cite}
\usepackage[cmex10]{amsmath} 
                             \usepackage{epsf}
\usepackage{xcolor}

\usepackage{amssymb}
\usepackage{epsfig,verbatim}
\usepackage{algorithm}
\usepackage{graphicx}
\usepackage{tikz}
\usepackage{pgfplots}
\usepackage{tkz-euclide}
\usepackage{tikz-3dplot}

\usepackage{algpseudocode}

\usepackage{hyperref}

\long\def\comment#1{}

\newfont{\bbb}{msbm10 scaled 700}

\newfont{\bb}{msbm10 scaled 1100}

\newcommand{\bv}{{\bf b}}

\newcommand{\uv}{{\bf u}}

\newcommand{\vv}{{\bf v}}
\newcommand{\xv}{{\bf x}}
\newcommand{\yv}{{\bf y}}

\newcommand{\Nm}{{\bf N}}

\newcommand{\Um}{{\bf U}}
\newcommand{\Wm}{{\bf W}}
\newcommand{\Vm}{{\bf V}}
\newcommand{\Xm}{{\bf X}}
\newcommand{\Ym}{{\bf Y}}
\newcommand{\Zm}{{\bf Z}}

\newcommand{\Ac}{{\cal A}}
\newcommand{\Bc}{{\cal B}}

\newcommand{\Ec}{{\cal E}}

\newcommand{\Hc}{{\cal H}}
\newcommand{\Ic}{{\cal I}}
\newcommand{\Jc}{{\cal J}}

\newcommand{\Nc}{{\cal N}}

\newcommand{\Pc}{{\cal P}}

\newcommand{\Sc}{{\cal S}}

\renewcommand{\det}{{\hbox{det}}}

\usepackage{times}
\usepackage{amsthm}

\newtheorem{theorem}{Theorem}

\newtheorem{lemma}{Lemma}

\newtheorem{corollary}{Corollary}

\newtheorem{remark}{Remark}

\newcommand{\argmin}{\operatornamewithlimits{argmin}}

\begin{document}
\title{Retrieving Data Permutations from Noisy Observations: High and Low Noise Asymptotics} 

\author{
\IEEEauthorblockN{Minoh Jeong$^\dagger$, Alex Dytso$^{\star}$,  Martina Cardone$^\dagger$}
$^\dagger$ University of Minnesota, Minneapolis, MN 55455, USA, Email: \{jeong316, mcardone\}@umn.edu\\
$^{\star}$ New Jersey Institute of Technology, Newark, NJ 07102, USA Email: alex.dytso@njit.edu
\thanks{The work of M. Jeong and M. Cardone was supported in part by the U.S. National Science Foundation under Grant CCF-1849757. 
}
}

\maketitle

\begin{abstract}
This paper considers the problem of recovering the permutation of an $n$-dimensional random vector $\Xm$  observed in Gaussian noise.  
First, a general expression for the probability of error is derived when a linear decoder (i.e., linear estimator followed by a sorting operation) is used. The derived expression holds with minimal assumptions on the distribution of $\Xm$ and when the noise has memory.  
Second,  for the case of isotropic noise (i.e., noise with a diagonal scalar covariance matrix), the rates of convergence of the probability of error are characterized in the high and low noise regimes.
 In the low noise regime, for every dimension $n$, the probability of error is shown to behave proportionally to $\sigma$, where $\sigma$ is the noise standard deviation.  Moreover, the slope is computed exactly for several distributions and it is shown to behave quadratically in $n$.
In the high noise regime,  for every dimension $n$, the probability of correctness is shown to behave as $1/\sigma$, and the exact expression for the rate of convergence is also provided.
\end{abstract}

\section{Introduction}
The problem of recovering data permutations from noisy observations is becoming a common task of modern communication and computing systems. For example, systems based on data sorting operations, such as a recommender system or a data analysis system, make use of the data permutations and leverage the information that can be obtained from the data ordering. In particular, recommender systems clearly utilize the sorting information in order to optimize their next recommendation. As for the case of a recommender system, data analysis systems are also often interested in rankings of massive data sets rather than in the exact values of the data. In such systems, users may desire to enclose their data when it contains sensitive information. A common solution to privatize individual data  is to add a sufficient amount of random noise to guarantee the desired privacy level \cite{dwork2008differential}.  However, adding too much noise can render the task of recovering a permutation impossible as the data will be too noisy. Therefore,  for a given noise level,  it is important to understand the fundamental limits of the data permutation recovery problem.

In this work, following preliminary works in~\cite{jeong20} and~\cite{ourJSAIT}, we study the data permutation recovery problem in the framework of an $M$-ary hypothesis testing.   
The specific goal of this paper is to study fundamental limits of such problem under the constraint that a {\em linear decoder} (i.e., linear estimator followed by a sorting operation) is employed.  
Studying linear decoders is interesting for several reasons. First, as it was shown in~\cite{jeong20} linear decoders are optimal (i.e., they lead to the smallest probability of error) when the noise is isotropic, and the distribution of the input data is exchangeable. Second, the optimal decoder can be linear even if the noise is colored; see~\cite{ourJSAIT} for the exact conditions. Third, linear decoders have at most polynomial complexity in the data dimension and hence, they are suitable for practical implementations. 

The structure of the paper is as follows. In Section~\ref{sec:PrForm}, we introduce the notation and  formally define the problem.  
In Section~\ref{sec:Prob_Error_Charac}, we characterize the probability of error when linear decoders are used. The derived expression holds with minimal assumptions on the distribution of the data  and holds when the noise has memory.   In Section~\ref{sec:Isotropic_Noise}, we utilize the expression for the error probability derived  in Section~\ref{sec:Prob_Error_Charac} and  characterize the {\em asymptotic} behavior of the probability of error for the isotropic noise case (i.e., when the noise covariance matrix is a diagonal scalar matrix) in the low and high noise regimes.  For example, we show that the probability of error linearly increases in $\sigma$ (i.e., the standard deviation of the noise) in the low noise regime (i.e., when $\sigma \to 0$). We derive the exact slope and we show it to be at most a quadratic function of the data dimension for a general class of distributions. In addition, we show that the behavior of the probability of correctness in the high noise regime (i.e., when $\sigma \to \infty$) is proportional to $\frac{1}{\sigma}$, and we characterize the exact slope.

\subsection{Related work}

Permutation associated estimation problems have recently gained significant importance and are studied in various fields~\cite{searle1973prediction,portnoy1982maximizing,nomakuchi1988lemma,nomakuchi1988,Collier,Pananjady2018,Pananjady2017,rigollet2019uncoupled,Unnikrishnan2018,Haghighatshoar2018_2,Zhang2019,dokmanic2019permutations,tsakiris19a,tsakiris2018eigenspace,Dytso2019}. 
The ranking (e.g., data permutation) estimation problem under a joint Gaussian distribution was investigated in~\cite{searle1973prediction,portnoy1982maximizing,nomakuchi1988lemma,nomakuchi1988}. In particular, in~\cite{searle1973prediction} the author considered a pairwise ordering for the bivariate case; the extended version to the $n$-dimension was considered in~\cite{portnoy1982maximizing}. The generalization of the assumption of a Gaussian distribution to an elliptically contoured distribution can be found in~\cite{nomakuchi1988lemma,nomakuchi1988}. The authors in~\cite{searle1973prediction,portnoy1982maximizing,nomakuchi1988lemma,nomakuchi1988} analyzed the structure of the covariance matrix that maximizes the probability of correctness of such estimation problems using the minimum mean square error (MMSE) estimator. In~\cite{ourJSAIT}, the MMSE estimator was shown to be the only linear estimator that achieves the minimum probability of error for the ranking estimation problem.
Most of recent works study a problem based on a linear regression framework premultiplied by an unknown permutation matrix, which suitably models the problem with unknown labels. In~\cite{Collier}, the feature matching problem in computer vision was formulated as a permutation recovery problem. The multivariate linear regression model with an unknown permutation was studied in~\cite{Pananjady2017,Pananjady2018}. 
The authors provided necessary and sufficient conditions on the signal-to-noise ratio for an exact permutation recovery and characterized the minimax prediction error.
The isotonic regression without data labels, namely the {\em uncoupled isotonic regression}, was discussed in~\cite{rigollet2019uncoupled}. 
Data estimation given randomly selected measurements -- referred to as {\em unlabeled sensing} -- was studied in \cite{Unnikrishnan2018,Haghighatshoar2018_2,Zhang2019}. 
In~\cite{Unnikrishnan2018}, the authors characterized a necessary condition on the dimension of the observation vector for uniquely recovering the original data in the noiseless case.
A generalized framework of unlabeled sensing was presented in~\cite{dokmanic2019permutations,tsakiris19a,tsakiris2018eigenspace}.
The estimation of a sorted vector based on noisy observations was proposed in~\cite{Dytso2019}, where the MMSE estimator on sorted data was characterized as a linear combination of estimators on the unsorted data. 

\section{Notation and Framework}
\label{sec:PrForm}
\noindent{\bf{Notation.}} 
Boldface upper case letters $\mathbf{X}$ denote vector random variables; the boldface lower case letter $\mathbf{x}$ indicates a specific realization of $\mathbf{X}$; 
$X_{i:n}$ denotes the $i$-th order statistics of $\Xm$;
$\|\Xm\|$ is the norm of $\Xm$;
$[n_1: n_2]$ is the set of integers from $n_1$ to $n_2 \geq n_1$;
$I_n$ is the identity matrix of dimension $n$; 
$\mathbf{0}_n$ is the column vector of dimension $n$ of all zeros; 
calligraphic letters indicate sets;
$|\mathcal{A}|$ is the cardinality of $\mathcal{A}$; for $\mathcal{A}$ and $\mathcal{B}$, $\mathcal{A} \setminus \mathcal{B}$ is the set of elements that belong to $\mathcal{A}$ but not to $\mathcal{B}$, $\mathcal{A} \cap \mathcal{B}$ is the set of elements that belong both to $\mathcal{A}$ and $\mathcal{B}$, and $\mathcal{A} \cup \mathcal{B}$ is the set of elements which are in either set.
For a set $\Sc \subseteq \mathbb{R}^{n}$, ${\rm Vol}(\Sc)$ denotes the volume, i.e., the $n$-dimensional Lebesgue measure.
For two $n$-dimensional vectors ${\bf x}$ and ${\bf y}$,  if for all $i \in[1:n]$, the $i$-th element of ${\bf x}$ is larger than or equal to the $i$-th element of ${\bf y}$, then we use ${\bf x} \ge {\bf y}$. 
Finally, the multiplication of a matrix $A$ by a set $\Bc$ is denoted and defined as $A\Bc=\{A\xv:\xv\in\Bc\}$.

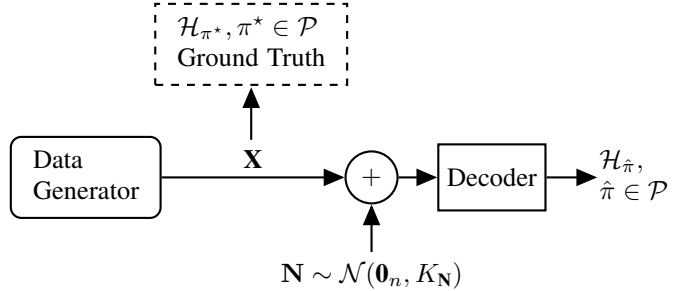
\begin{figure}
        \centering

        \tikzset{%
  block/.style    = {draw, thick, rectangle, minimum height = 2.5em,
    minimum width = 2.5em},
  sum/.style      = {draw, circle, node distance = 1.8cm}, 
  input/.style    = {coordinate}, 
  output/.style   = {coordinate} 
  
   ret5/.style      = {draw=gray,dashed}, 
}

\newcommand{\suma}{\Large$+$}
\newcommand{\inte}{$\displaystyle \int$}
\newcommand{\derv}{\huge$\frac{d}{dt}$}

\begin{tikzpicture}[auto, thick, node distance=1.3cm, >=triangle 45]
width=6cm,
height=5.7cm,
\draw 
         node[block,rounded corners] (input1) {$ \begin{array}{l} \text{Data}  \\  \text{Generator} \end{array} $}
         node[sum,  right=2cm,right of=input1] (Q1) {$+$}
         node [ below of=Q1] (ret2) {  $\mathbf{N}   \sim  \mathcal{N}(\textbf{0}_n,K_{\textbf{N}})$}
         node [block, right=0.3cm,  right of=Q1] (ret3) {Decoder}
         node  [ right=6.8cm](ret4){  \hspace{-0.4cm}  $ \begin{array}{l}  \mathcal{H}_{\hat{\pi}}, \\  \hat{\pi} \in \mathcal{P} \end{array}$ }
            node [ block,dashed, right=2.2cm, yshift=0.47cm,  above of=input1] (ret5){  $ \begin{array}{l} \mathcal{H}_{\pi^\star},   \pi^\star \in \mathcal{P}  \\  \text{Ground Truth} \end{array} $   }
          
         ;
         
	\draw[->](input1) -- node {$\textbf{X} \text{}$}(Q1);
	 \draw[->](ret2) --node {} (Q1);
	  \draw[->](Q1) --node {} (ret3);
	    \draw[->](ret3) --node {} (ret4);
	    
	     \draw[->](2.2,0.5) --node {} (2.2,1.2);
\end{tikzpicture}
        \caption{Graphical representation of the considered framework.}
        \label{fig:Framework}
    \end{figure}

\smallskip
We consider the framework in Fig.~\ref{fig:Framework}, where an $n$-dimensional random vector $\mathbf{X} \in \mathbb{R}^n$ is first generated according to a certain distribution and then passed through an additive noisy channel with Gaussian transition probability, the output of which is denoted as $\mathbf{Y}$. 
Thus, we have
$\mathbf{Y} = \mathbf{X} + \mathbf{N}$, with $\mathbf{N} \sim \mathcal{N}(\mathbf{0}_n,K_{\mathbf{N}})$ where $K_{\mathbf{N}}$ is the covariance matrix of the additive noise $\mathbf{N}$, and where $\mathbf{X}$ and $\mathbf{N}$ are independent.

In this work, we are interested in studying the {\em probability of error} of the ``data permutation recovery'' problem formulated in~\cite{jeong20,ourJSAIT} that, given the observation of $\mathbf{Y}$, seeks to retrieve the permutation (among the $n!$ possible ones) according to which the vector $\mathbf{X}$ is sorted.
Specifically, this problem can be formulated within a hypothesis testing framework with $n!$ hypotheses $\mathcal{H}_{\pi}, \pi \in \mathcal{P}$, where $\mathcal{P}$ is the collection of all permutations of the elements of $[1:n]$, and where $\mathcal{H}_{\pi}$ is the hypothesis that $\mathbf{X}$ is an $n$-dimensional vector sorted according to the permutation $\pi \in \mathcal{P}$, that is 
\begin{align}
\label{eq:HR}
\Hc_\pi = \{\mathbf{x} \in\mathbb{R}^n :x_{\pi_{1}} \le x_{\pi_2} \le \cdots \le x_{\pi_n} \},
\end{align} 
with $x_{\pi_i}, i \in [1:n]$ being the $\pi_i$-th element of $\mathbf{x}$, and $\pi_i, i \in[1:n]$ being the $i$-th element of $\pi$.
Given this, the {\em optimal} decoder in Fig.~\ref{fig:Framework} will output $\mathcal{H}_{\hat{\pi}}, \hat{\pi} \in \mathcal{P}$ such that
\begin{align}
\label{eq:OptCrit}
\mathcal{H}_{\hat{\pi}}: \  \hat{\pi} = \argmin_{\pi \in \mathcal{P}} \ \{ \Pr \left (\mathcal{H}_{\pi} \neq \mathcal{H}_{\pi^\star}\right ) \},
\end{align}
where $\pi^\star$ denotes the permutation according to which the random vector $\mathbf{X}$ is sorted.
In particular, the decoder will declare that the input vector $\mathbf{x} \in \mathcal{H}_{\pi}$ if and only if the observation vector $\mathbf{y} \in \mathcal{R}_{\pi,K_{\mathbf{N}}}$, where $\mathcal{R}_{\pi,K_{\mathbf{N}}}, \pi \in \mathcal{P}$ are the so-called {\em optimal} decision regions\footnote{The notation $\mathcal{R}_{\pi,K_{\mathbf{N}}}$ indicates that, in general, the decision regions  might be functions of the noise covariance matrix $K_{\mathbf{N}}$.}, which can be derived by leveraging the maximum a posterior probability (MAP) criterion~\cite[Appendix~3C]{Kay1998} and are given by~\cite{jeong20,ourJSAIT}
\begin{align}\label{region}
&\mathcal{R}_{\pi,K_{\mathbf{N}}}
=\left\{\yv \in \mathbb{R}^n:f_\Ym(\yv,\mathcal{H}_{\pi})>\max_{\substack{\tau \in \mathcal{P} \\ \tau \neq \pi}}f_\Ym(\yv,\mathcal{H}_{\tau})\right\},
\end{align} 
where $f_\Ym(\yv,\mathcal{H}_{\pi}) = f_\Ym(\yv|\mathcal{H}_{\pi})\Pr (\mathcal{H}_{\pi})$ with $f_\Ym(\yv|\mathcal{H}_{\pi})$ denoting the conditional probability density function of $\Ym$ given that $\Xm \in \mathcal{H}_{\pi}$.
In order to guarantee that the collection $\{\mathcal{R}_{\pi,K_{\mathbf{N}}}, \pi \in \mathcal{P}\}$ is a partition of the $n$-dimensional space, we assume that if $\mathbf{y} \in \{ \mathcal{R}_{\pi,K_{\mathbf{N}}}, \pi \in \mathcal{S}, \mathcal{S} \subseteq \mathcal{P}, |\mathcal{S}|>1\}$, then one of the hypotheses $\mathcal{H}_{\pi}, \pi \in \mathcal{S}$ is arbitrarily selected.

\section{Probability of Error with Linear Decoder}
\label{sec:Prob_Error_Charac}
In this section, we focus on characterizing the probability of error of the data permutation recovery problem introduced in Section~\ref{sec:PrForm}. Given the hypothesis and decision regions defined in~\eqref{eq:HR} and~\eqref{region}, we have that the error probability $P_e$ is given~by\begin{subequations}
\label{eq:ErrorProbGen}
\begin{align}
P_e &= 1 - P_c,
\\
P_c &=   \sum_{\pi\in\Pc} \Pr\left(  \{ {\bf Y} \in \mathcal{R}_{\pi,K_{\mathbf{N}}} \}   \cap   \{ {\bf X} \in \mathcal{H}_\pi  \}   \right),
\end{align}
\end{subequations}
where $P_c$ is the probability of correctness.

In particular, we assess the probability of error when a {\em linear} decoder is employed.
This decoder first computes a permutation-independent linear transformation $\mathbf{y}_{\ell}$ of $\mathbf{y}$, i.e., $\mathbf{y}_{\ell} = A \mathbf{y} + \mathbf{b}$, where $A \in \mathbb{R}^{n \times n}$ and $\mathbf{b} \in \mathbb{R}^n$ are the same for all permutations, and then it outputs the permutation according to which $\mathbf{y}_{\ell}$ is sorted. The decision regions in~\eqref{region} when a linear decoder is used become
\begin{align}
\label{eq:regionLinear}
\mathcal{R}_{\pi,K_{\mathbf{N}}} = A \mathcal{H}_{\pi} + \mathbf{b}, \ \forall \pi \in \mathcal{P}.
\end{align}
Our choice of assessing the probability of error performance of a linear decoder stems primarily from its low complexity (at most polynomial in $n$) compared to a brute force evaluation of the optimal test~\eqref{region}, which has a practically prohibitive complexity of $n!$.
Moreover, for the case $\Xm \sim \mathcal{N} (\mathbf{0}_n, I_n)$ it has been shown in~\cite{ourJSAIT} that a linear decoder is indeed optimal, i.e., it minimizes the probability of error, under certain conditions on the noise covariance matrix $K_\Nm$.

We next derive an expression for the probability of error when a linear decoder is used. Towards this end, for each $\pi \in \mathcal{P}$, we define a matrix $T_{\pi} \in \mathbb{R}^{(n-1) \times n}$ such that
\begin{align}
\label{eq:Tmatr}
(T_{\pi})_{i,j} = 1_{\{j = \pi_{i+1}\}} - 1_{\{j = \pi_i\}},
\end{align}
where $1_{\{x=y\}} = 1$ if and only if $x=y$ and is equal to zero otherwise.
For instance, let $n=4$ and consider $\pi = \{4,2,1,3\}$; then, we have that
\begin{align*}
T_{\{4,2,1,3\}} =
\begin{bmatrix}
0 & 1 & 0 & -1
\\
1 & -1 & 0 & 0
\\
-1 & 0 & 1 & 0
\end{bmatrix}.
\end{align*}
The theorem below provides an expression for the error probability of the data permutation recovery problem when a linear decoder is used.
\begin{theorem}\label{thm:PcLE}
Let $\Xm$ be an exchangeable random vector\footnote{A sequence of random variables $U_1, … , U_n$ is said to be exchangeable if, for any permutation $(\pi_1, … , \pi_n)$ of the indices $[1:n]$, we have that $(U_1, …, U_n)$ is equal in distribution to $(U_{\pi_1},  …, U_{\pi_n})$.}.
Then, for any invertible $A$ and $\mathbf{b}$ defined in~\eqref{eq:regionLinear} and any noise covariance matrix $K_\Nm$, the probability of error is given by
\begin{align}
\label{eq:PeGeneral}
	P_e \!=\! 1 \!-\! \frac{1}{n!}  \sum_{\pi\in\Pc} \mathbb{E}\left[ Q_{\tilde{K}_{\pi}} \! \left( - T_\pi A^{-1}(\Xm \!-\!\bv) \right)  \;\middle | \;    {\bf X} \in \mathcal{H}_\pi  \right],
\end{align}
where $\tilde{K}_{\pi}=T_\pi A^{-1}K_\Nm A^{-T}T_\pi^T \in \mathbb{R}^{(n-1) \times (n-1)}$ with $T_{\pi}, \pi \in \mathcal{P}$ given by~\eqref{eq:Tmatr}, and where $Q_{\tilde{K}_{\pi}}\!(\cdot )\!$ is the multivariate Gaussian Q-function with covariance $\tilde{K}_{\pi}$.
\end{theorem}
\begin{IEEEproof}
By substituting the decision regions in~\eqref{eq:regionLinear} inside~\eqref{eq:ErrorProbGen} and by using the Bayes' theorem, we obtain
\begin{align}\label{eq:PcLE}
&P_c  =   \sum_{\pi\in\Pc} \Pr\left( {\bf Y} \in A \mathcal{H}_\pi  +\bv \;\middle | \;    {\bf X} \in \mathcal{H}_\pi     \right)   \Pr\left(     {\bf X} \in \mathcal{H}_\pi     \right) \nonumber\\
&{\overset{\rm (a)}{=}}   \frac{1}{n!}\! \sum_{\pi\in\Pc} \Pr\left( \Xm + K_{\Nm}^{\frac{1}{2}}\Zm -\bv \in A \mathcal{H}_\pi   \! \;\middle | \;   \! {\bf X} \in \mathcal{H}_\pi     \right)    \nonumber\\
&{\overset{\rm (b)}{=}}   \frac{1}{n!} \! \sum_{\pi\in\Pc} \mathbb{E}\!\left[ \Pr\!\left( \Xm \!+\!  K_{\Nm}^{\frac{1}{2}}\Zm - \bv \in \! A \mathcal{H}_\pi  \!  \;\middle | \;   \! {\bf X}     \right)  \!   \;\middle | \;  \!  {\bf X} \!\in\! \mathcal{H}_\pi  \right],\! 
\end{align}
where $\rm (a)$ follows from the fact that $\Xm$ is exchangeable and hence, $\Pr(\Xm\in\Hc_\pi) = \frac{1}{n!},~\forall \pi\in\Pc$ and letting $\Zm  \sim \mathcal{N} (\mathbf{0}_n, I_n)$, and $\rm (b)$ is due to the law of total expectation.

We now focus on the conditional probability inside the conditional expectation in~\eqref{eq:PcLE}. For each $\Hc_\pi,~\forall\pi\in\Pc$ we have
\begin{align}\label{eq:probHi}
	&  \Pr\left( {\bf X}+K_\Nm^{\frac{1}{2}} {\bf Z} -\bv  \in A \mathcal{H}_\pi  \;\middle | \; {\bf X}  \right) \nonumber\\
	& = \Pr\left( A^{-1}({\bf X} - \bv)+ A^{-1} K_\Nm^{\frac{1}{2}} {\bf Z}  \in \mathcal{H}_\pi  \;\middle | \; {\bf X}  \right) \nonumber\\
	& = \Pr\left( A^{-1}({\bf X} - \bv) + \Um  \in \mathcal{H}_\pi  \;\middle | \; {\bf X}  \right), 
\end{align}
where the last equality follows by letting $\Um = A^{-1} K_\Nm^{\frac{1}{2}} \Zm$. 
Note that $\Um\sim\Nc(\mathbf{0}_n, A^{-1}K_\Nm A^{-T})$.

Then, given $\Xm$, the event inside the conditional probability in~\eqref{eq:probHi} can be expressed as
\begin{align}  
	& \left\{A^{-1}({\bf X} - \bv) + \Um  \in \mathcal{H}_\pi \right\} \nonumber \\
	& =  \bigcap_{k=1}^{n-1} \! \left \{  \left(A^{-1}(\Xm \!-\! \bv)\right)_{\!\pi_{k}} \!\!-\! \left(A^{-1}(\Xm \!-\! \bv)\right)_{\!\pi_{k+1}} \! \le  U_{\pi_{k+1}} \!-\! U_{\pi_k}    \right \} \nonumber \\
	& = \left \{ - T_\pi A^{-1}(\Xm-\bv)   \le  T_\pi \Um \right \},
\end{align}
where the last equality follows by using the definition of $T_{\pi}, \pi \in \mathcal{P}$ in~\eqref{eq:Tmatr}.
By introducing a random vector $\Vm_{\pi} = T_{\pi}\Um\sim\Nc(\mathbf{0}_{n-1},\tilde{K}_{\pi})$, where $\tilde{K}_{\pi} = T_\pi A^{-1}K_\Nm A^{-T}T_\pi^T$, we have an equivalent expression for~\eqref{eq:probHi} as $\Pr\left( {\bf X}+K_\Nm^{\frac{1}{2}} {\bf Z} -\bv  \in A \mathcal{H}_\pi  \;\middle | \; {\bf X}  \right) = \Pr\left(    - T_\pi A^{-1}(\Xm-\bv) \le  \Vm_\pi   \;\middle | \; {\bf X} \right  )$.
By substituting this inside~\eqref{eq:PcLE}, we obtain
\begin{align}
	P_c 
	& = \!  \frac{1}{n!}  \!\sum_{\pi\in\Pc} \!\mathbb{E}\left[ \Pr\left(  \!- T_\pi A^{-1}(\Xm-\bv)  \le  \Vm_\pi    \;\middle | \; {\bf X} \right  )   \;\middle | \;    {\bf X} \in \mathcal{H}_\pi  \right]\nonumber\\
	& =  \frac{1}{n!}  \sum_{\pi\in\Pc} \mathbb{E}\left[   Q_{\tilde{K}_{\pi}} \left( -T_\pi A^{-1}(\Xm-\bv) \right)  \;\middle | \;    {\bf X} \in \mathcal{H}_\pi  \right],
\end{align}
where the last equality follows by letting $Q_{\tilde{K}_{\pi}}(\cdot)$ be the multivariate Gaussian Q-function with covariance $Q_{\tilde{K}_{\pi}}$.
We conclude the proof of Theorem~\ref{thm:PcLE} by using $P_e = 1 - P_c$.
\end{IEEEproof}
We highlight that~\eqref{eq:PeGeneral} holds with minimal assumption on the distribution of $\Xm$ (i.e., exchangeability) and hence, it can be used to study the error probability of the data permutation recovery problem in various noise settings, e.g., noise has memory, noise is isotropic.
In the remaining of this paper, we will focus on the isotropic noise scenario, i.e., we assume that $K_\Nm$ is a diagonal scalar matrix.

\section{Isotropic Noise}
\label{sec:Isotropic_Noise}
We here study the error probability of the data permutation recovery problem when the noise is isotropic, i.e., $K_\Nm = \sigma^2 I_n$.
Under this assumption, the regions $\mathcal{R}_{\pi,K_{\mathbf{N}}}, \pi \in \mathcal{P}$ in~\eqref{eq:regionLinear} depends on $K_\Nm$ only through the parameter $\sigma$ and hence, we let $\mathcal{R}_{\pi,K_{\mathbf{N}}} = \mathcal{R}_{\pi,\sigma}$.
Moreover, when $\Xm$ is exchangeable, it has been shown in~\cite{jeong20} that $\mathcal{R}_{\pi,\sigma} = \mathcal{H}_{\pi},\pi \in \mathcal{P}$, i.e., for the isotropic noise setting the optimal decoder is indeed linear and hence, the probability of error in Theorem~\ref{thm:PcLE} is the minimum.

In Section~\ref{sec:ErrProbIso}, we will evaluate the probability of error $P_e$ in~\eqref{eq:PeGeneral} when $K_\Nm = \sigma^2 I_n$ and then in Section~\ref{sec:ErrProbIsoLN} and Section~\ref{sec:ErrProbIsoHN}, we will use this expression to derive the rates of convergence of $P_e$ in the low noise regime (i.e., $\sigma \to 0$) and high noise regime (i.e., $\sigma \to \infty$), respectively.

\subsection{Probability of Error}
\label{sec:ErrProbIso}
Under the assumption $K_\Nm = \sigma^2 I_n$ we have that $\mathcal{R}_{\pi,\sigma} = \mathcal{H}_{\pi},\pi \in \mathcal{P}$~\cite{jeong20} and hence, with reference to~\eqref{eq:regionLinear}, we have that $A=I_n$ and $\mathbf{b}=\mathbf{0}_n$. 
Moreover, by substituting these values inside $\tilde{K}_{\pi} \in \mathbb{R}^{(n-1) \times (n-1)}$ in Theorem~\ref{thm:PcLE}, we obtain
\begin{align}
\label{eq:KTilde}
\begin{split}
\tilde{K}_{\pi}&=T_\pi A^{-1}K_\Nm A^{-T}T_\pi^T  = \sigma^2 T_\pi  T_\pi^T = \sigma^2 \tilde{K},
\\ (\tilde{K})_{i,j} &= \left \{
\begin{array}{ll}
2 & i=j,
\\
-1 & i = j+1 \ \text{and} \ j=i+1,
\\
0 & \text{otherwise,}
\end{array}
\right . 
\end{split}
\end{align}
that is $\tilde{K} \in \mathbb{R}^{(n-1)\times (n-1)}$ is a tridiagonal Toeplitz matrix.

The probability of error in the isotropic noise scenario is then given by the next corollary.
\begin{corollary}\label{prop:Peiso}
Let $\Xm$ be an exchangeable random vector and let $K_\Nm = \sigma^2 I_n$.
Then, for an arbitrary $\pi \in \mathcal{P}$, the probability of error is given by
\begin{align}
	P_e = 1 - \mathbb{E}\left[ Q_{\sigma^2\tilde{K}} \! \left( -  T_\pi \Xm \right)  \;\middle | \;    {\bf X} \in \mathcal{H}_\pi  \right],
\end{align}
where $\tilde{K}$ is defined in~\eqref{eq:KTilde} and where $Q_{\tilde{K}}(\cdot)$ is the multivariate Gaussian Q-function with covariance $\sigma^2 \tilde{K}$.
\end{corollary}
\begin{IEEEproof}
By substituting the expression of $\sigma^2 \tilde{K}_{\pi}$ in~\eqref{eq:KTilde} inside~\eqref{eq:PeGeneral}, we obtain
\begin{align}\label{eq:Peisopf2}
	P_e =1 -  \frac{1}{n!}  \sum_{\pi\in\Pc} \mathbb{E}\left[ Q_{\sigma^2 \tilde{K}}  \left(-T_\pi \Xm \right)  \;\middle | \;    {\bf X} \in \mathcal{H}_\pi  \right].
\end{align}
We note that $\sigma^2 \tilde{K}$ and the distribution of $T_\pi \Xm \ |\ \Xm\in\Hc_\pi$ are independent of $\pi \in \mathcal{P}$ and hence, the conditional expectation in~\eqref{eq:Peisopf2} is constant in $\pi \in \mathcal{P}$. Since $|\mathcal{P}|=n!$, we obtain 
\begin{align}\label{eq:Peisopf3}
	P_e 
	& =  1 - \mathbb{E}\left[ Q_{\sigma^2\tilde{K}} \! \left( -  T_\tau \Xm \right)  \;\middle | \;    {\bf X} \in \mathcal{H}_\tau  \right],
\end{align}
where $\tau\in\Pc$ can be arbitrary.
\end{IEEEproof}
We note that~\eqref{eq:Peisopf2} is a function of $\sigma$ and hence, in what follows we will use $P_e(\sigma)$ to highlight this dependence.

\subsection{Low Noise Asymptotic}
\label{sec:ErrProbIsoLN}
We here focus on the asymptotic behavior of the probability of error in the low noise regime (i.e., $\sigma \to 0$).  
In particular, the next result, proved in Appendix~\ref{app:LowNoise}, shows that the probability of error in this regime is approximately linear in $\sigma$. 
\begin{theorem}\label{prop:sigto0}
Let $\Xm$ consist of $n$ i.i.d. random variables generated according to $X$.  Let $X'$ be an independent copy of $X$ and assume that
\begin{equation}
 f_{X-X^\prime}(x)< \infty,~\forall x \in \mathbb{R}. 
\end{equation} 
Then,
\begin{equation}\label{eq:lims0}
	\lim_{\sigma\to0} \frac{P_e(\sigma)}{\sigma}
	=   \sum_{i=1}^{n-1} \frac{f_{W_i}\! \left( 0^+ \right)}{\sqrt{\pi}},
\end{equation}
where   $W_i = X_{i+1:n}  - X_{i:n},~i\in[1:n-1]$.
\end{theorem}
\begin{remark}\label{re:exchange}
The   i.i.d. assumption on $\Xm$  in Theorem~\ref{prop:sigto0} can be relaxed to the case of exchangeable $\Xm$,  provided that the following holds: for $1\le i < j \le n-1$,\begin{subequations}\label{eq:exchjoint}
\begin{align}
	& f_{W_i}(u)<\infty,~\forall u\in\mathbb{R}_+, \label{eq:exchjoint1}\\
	& f_{W_i,W_j}(u,v)<\infty,~\forall(u,v)\in\mathbb{R}_+^2.
\end{align}
\end{subequations}
Then, under these conditions, for the case of an exchangeable $\Xm$ we have the same result as in Theorem~\ref{prop:sigto0}.
\end{remark}
\begin{remark}
The quantity $f_{W_i} \left( 0^+ \right)$ in~\eqref{eq:lims0} can be computed as follows~\cite{spacingPyke}
\begin{align}
	f_{W_i}(0^+) 
	 & = \frac{n! \! \int_{-\infty}^{\infty} F(x)^{i-1} \!\left( 1\! - \! F(x) \right)^{n-i-1} \! f^2(x)\ {\rm d} x}{ (i-1)! (n-i-1)!} \nonumber
\\& = \frac{n!  \mathbb{E} \left[ U^{i-1} \left( 1 - U \right)^{n-i-1} f(F^{-1}(U)) \right] }{ (i-1)! (n-i-1)!},
\label{eq:fWi}
\end{align}
where the last step uses the probability integral transformation theorem and the quantile function theorem~\cite{PIT} with $U\sim \text{Unif}(0,1)$.
\end{remark}
We now show that, under the condition $\sup_{x \in \mathbb{R}} f_X(x)=c$, the asymptotic behavior of the probability of error in the low noise regime for an i.i.d. $\Xm$ is upper bounded by $O(n^2)$. In particular, we have the following lemma.
\begin{lemma}
\label{eq:UBLN}
Assume that $\sup_{x \in \mathbb{R}} f_X(x)=c$, where $c \in \mathbb{R}$ is a constant. Then,
\begin{equation}
\lim_{\sigma\to0} \frac{P_e(\sigma)}{\sigma} \le c  \frac{n(n-1)}{\sqrt{\pi}}. 
\end{equation} 
\end{lemma}
\begin{IEEEproof}
By using the expression in~\eqref{eq:fWi}, we have that
\begin{align*}
	f_{W_i}(0^+) 
	& = \frac{n!  \mathbb{E} \left[ U^{i-1} \left( 1 - U \right)^{n-i-1} f(F^{-1}(U)) \right] }{ (i-1)! (n-i-1)!}\\
	& \le c  \frac{n!  \mathbb{E} \left[ U^{i-1} \left( 1 - U \right)^{n-i-1}  \right] }{ (i-1)! (n-i-1)!}\\
& = c\frac{n! \int_{0}^{1} x^{i-1} \left(  1-x \right)^{n-i-1}  \ {\rm d} x}{(i-1)! (n-i-1)!}  \nonumber \\
	& = c \frac{n!  }{(i-1)! (n-i-1)! } \frac{  \Gamma(n-i) \Gamma(i) }{\Gamma(n) }  \nonumber \\
	& = cn,
\end{align*}
where $\Gamma (\cdot)$ is the gamma function and where the inequality follows by using the bound  $f(F^{-1}(U)) \le c =\sup_{x \in \mathbb{R}} f(x) $.
Hence, \eqref{eq:lims0} can be upper bounded as
\begin{equation}
	\lim_{\sigma\to0} \frac{P_e(\sigma)}{\sigma} = \sum_{i=1}^{n-1} \frac{f_{W_i}(0^+)}{\sqrt{\pi}}  
	 \le c \frac{ (n-1)n}{\sqrt{\pi}}.  \label{eq:General_upper_bound_last_step}
\end{equation}
This concludes the proof of Lemma~\ref{eq:UBLN}.
\end{IEEEproof}
We  conclude this section by providing some examples of~\eqref{eq:lims0} for a few distributions. 

\noindent
{\bf{Example~1.}} Consider $X\sim \text{Unif}(a,b ),~0\le a<b<\infty$. Then,
\begin{align}
\label{eq:SlopeUnif}
	& \lim_{\sigma\to0} \frac{P_e(\sigma)}{\sigma}
	 =  \frac{n (n-1)}{(b-a)\sqrt{\pi}} .
\end{align}
The proof of~\eqref{eq:SlopeUnif} can be found in Appendix~\ref{app:SlopeUnif}. \hfill $\diamond$

\noindent
{\bf{Example~2.}} Consider $X\sim \text{Exp} (\lambda ),~\lambda>0$. Then,
\begin{align}
\label{eq:SlopeExp}
	& \lim_{\sigma\to0} \frac{P_e(\sigma)}{\sigma}
	 = \frac{\lambda n (n-1)}{2\sqrt{\pi}}.
\end{align}
The proof of~\eqref{eq:SlopeExp} can be found in Appendix~\ref{app:SlopeExp}. \hfill $\diamond$

\noindent
{\bf{Example~3.}} Consider $X\sim  \Nc(0,1)$. Then,
\begin{align}
\label{eq:SlopeGauss}
	\frac{ \sqrt{2}  n(n-1)}{ 6 \pi  }
	& \le \lim_{\sigma\to0} \frac{P_e(\sigma)}{\sigma}
	 \le \frac{n(n-1)  }{\sqrt{2}\pi }.
\end{align}
Note that the upper bound in~\eqref{eq:SlopeGauss} follows from Lemma~\ref{eq:UBLN},
where we used the fact that  $c =\sup_{x \in \mathbb{R}} f(x) =\frac{1}{\sqrt{2\pi}}$.
For the lower bound we use the following inequality~\cite[Lemma~10.1]{boucheron2013concentration}:
\begin{align}\label{eq:exlowtech}
	f(F^{-1}(u)) 
	\ge  \sqrt{\frac{2}{\pi}} \min\{u,1-u\} 
	\ge \sqrt{\frac{2}{\pi}} u(1-u),
\end{align}
where the last step follows since $\min\{a,b\} \ge \frac{ab}{a+b},~\forall a>0,b>0$.
Combining the expression  in~\eqref{eq:fWi} and the bound in~\eqref{eq:exlowtech}, we arrive at the following lower bound,
\begin{align}
	f_{W_i}(0^+) 
	& \ge \sqrt{\frac{2}{\pi}} \frac{n!  \mathbb{E} \left[ U^{i} \left( 1 - U \right)^{n-i}  \right] }{ (i-1)! (n-i-1)!}\nonumber \\
	& = \sqrt{\frac{2}{\pi}} \frac{ i (n-i) }{ (n+1)  },
\end{align}
which implies the lower bound in~\eqref{eq:SlopeGauss}. \hfill $\diamond$

\subsection{High Noise Asymptotic}
\label{sec:ErrProbIsoHN}
We now focus on the asymptotic behavior of the probability of error in the high noise regime (i.e., $\sigma \to \infty$).   It is not difficult to argue that  if $\Xm$ is exchangeable, then we have that
\begin{equation}
\lim_{\sigma \to \infty} P_e (\sigma)=1-\frac{1}{n!}=P_e(\infty).  \label{eq:Limit_sigma_infinity}
\end{equation} 
The interpretation is that if $\sigma$ is large, then the output $\Ym$ carries no information of $\Xm$, and the decoder can only rely on the prior knowledge; hence, the best thing that the decoder can do is  to guess one of the $n!$ hypotheses. 

The next result, proved in Appendix~\ref{app:HighNoise}, sharpens the limit in \eqref{eq:Limit_sigma_infinity} by finding   the rate of convergence.

\begin{theorem} 
\label{thm:sigtoInf}
Let $\Xm$ be an exchangeable random vector such that $\mathbb{E} [ \|\Xm\| ]<\infty$. Then,
\begin{equation}\label{eq:liminf}
	\lim_{\sigma \to \infty} \frac{P_e(\infty) - P_e(\sigma)}{\frac{1}{\sigma}}
	 = \frac{1}{\sqrt{2 \pi}} \sum_{i=1}^{n-1} \alpha_i   \mathbb{E}  \left[  W_i \right],
\end{equation}
where    $W_i = X_{i+1:n}  - X_{i:n},~i\in[1:n-1]$ and
\begin{equation}\label{eq:alp}
	\alpha_i
	 =  \frac{ {\rm Vol}\left( \Ec (\mathbf{0}_{n-1},i) \cap  \Hc_{[1:n-1]} \right )}{{ \rm Vol} \left (\Bc(\mathbf{0}_{n-1},1) \right)},
\end{equation}
where $\Hc_{[1:n-1]}$ is defined in~\eqref{eq:HR}, $\Bc(\mathbf{0}_{n-1},1)$ is the $(n-1)$-dimensional ball centered at the origin with unitary radius, and $\Ec ( \mathbf{0}_{n-1},i)$ is the $(n-1)$-dimensional ellipsoid centered at the origin with unit radii along standard axes except a $\frac{1}{\sqrt{2}}$ radius along the $i$-th axis.
\end{theorem} 
Finding a closed-form expression for the $\alpha_i$'s in~\eqref{eq:alp} does not appear to be an easy task. In the next lemma, we provide upper and lower bounds on the $\alpha_i$'s, which lead to expressions that are amenable to computations.  
\begin{lemma}\label{lem:siginfBound}
In the high noise regime, the convergence rate of the probability of correctness can be bounded as
\begin{align*}
	 \frac{\mathbb{E}  \left[  R_n \right]}{\sqrt{\pi}(n-1)! 2^{\frac{n}{2}} }  
	\le \lim_{\sigma \to \infty} \frac{P_e(\infty) - P_e(\sigma)}{\frac{1}{\sigma}}
	 \le  \frac{ \mathbb{E}  \left[  R_n \right]}{\sqrt{2 \pi} (n-1)!}  ,
\end{align*}
where $R_n = X_{n:n} - X_{1:n}$.
\end{lemma}
\begin{IEEEproof}
We start by observing that
\begin{align}\label{eq:volsubset}
	\Bc \left( \mathbf{0}_{n-1},2^{-\frac{1}{2}} \right)
	\subset \Ec ( \mathbf{0}_{n-1},i)
	\subset \Bc \left( \mathbf{0}_{n-1},1\right),
\end{align}
that is, the ellipsoid $\Ec ( \mathbf{0}_{n-1},i)$: (i) contains the ball $\Bc \left( \mathbf{0}_{n-1},2^{-\frac{1}{2}} \right)$ since $\Ec ( \mathbf{0}_{n-1},i)$ has minimum radius equal to $2^{-\frac{1}{2}}$; and (ii) is contained inside the ball $\Bc \left( \mathbf{0}_{n-1},1\right)$ since $\Ec ( \mathbf{0}_{n-1},i)$ has maximum radius equal to $1$.

Thus, from~\eqref{eq:volsubset} we obtain
\begin{align}\label{eq:alpup}
	\alpha_i
	  \le  \frac{ {\rm Vol}\left( \Bc \left( \mathbf{0}_{n-1},1\right) \cap  \Hc_{[1:n-1]} \right )}{{ \rm Vol} \left (\Bc(\mathbf{0}_{n-1},1) \right)} 
	 = \frac{1}{(n-1)!},
\end{align}
where the last equality follows since $\Hc_{[1:n-1]}$ is a cone that occupies a $\frac{1}{(n-1)!}$ portion of the space and hence, ${\rm Vol}\left( \Bc \left( \mathbf{0}_{n-1},1\right) \cap \Hc_{[1:n-1]}\right) =  \frac{1}{(n-1)!}{\rm Vol}\left( \Bc \left( \mathbf{0}_{n-1},1\right) \right)  $.

Similarly, from~\eqref{eq:volsubset} we obtain
\begin{align}\label{eq:alplow}
	\alpha_i
	& \ge  \frac{ {\rm Vol}\left( \Bc \left( \mathbf{0}_{n-1}, 2^{-\frac{1}{2}} \right) \cap  \Hc_{[1:n-1]} \right )}{{ \rm Vol} \left (\Bc(\mathbf{0}_{n-1},1) \right)} \nonumber \\
	& = \left | \det \left (2^{-\frac{1}{2}} I_{n-1} \right ) \right | \frac{ {\rm Vol}\left( \Bc \left(  \mathbf{0}_{n-1}, 1 \right) \cap \Hc_{[1:n-1]} \right )}{{ \rm Vol} \left (\Bc(\mathbf{0}_{n-1},1) \right)} \nonumber \\
	& =  \frac{1}{2^{\frac{n-1}{2}} (n-1)!},
\end{align}
where in the equality  we have used the facts that: (i) $ 2^{\frac{1}{2}} I_{n-1} \Bc \left( \mathbf{0}_{n-1}, 2^{-\frac{1}{2}} \right) = \Bc \left( \mathbf{0}_{n-1}, 1 \right)$, (ii) $ 2^{\frac{1}{2}} I_{n-1}  \Hc_{[1:n-1]} =  \Hc_{[1:n-1]}$, and (iii) ${\rm Vol} (A\Sc ) = | \det(A) | {\rm Vol} (\Sc )$ for any invertible matrix $A$ and any set $\Sc$.

The proof of Lemma~\ref{lem:siginfBound} is concluded by substituting~\ref{eq:alpup} and~\eqref{eq:alplow} into~\eqref{eq:liminf} and by using the fact that
\begin{align}
	\sum_{i=1}^{n-1} \mathbb{E} [W_i] = \mathbb{E}[R_n],
\end{align}
where $R_n = X_{n:n} - X_{1:n}$ denotes the range~\cite{orderstat} of $\Xm$.
\end{IEEEproof}
We  conclude this section by providing some examples of the range $R_n$ for a few common distributions (see Appendix~\ref{app:ExamHighNoise} for the detailed computations). 
In particular, these examples show that the term $\frac{1}{(n-1)!}$ dominates in the expression of the rate for several distribution of interest.

\noindent
{\bf{Example~1.}} Consider $X\sim \text{Unif}(a,b ),~0\le a<b<\infty$. Then,
\begin{equation*}
	 \mathbb{E}[R_n]
	= (b-a)\frac{(n-1)}{n+1}.
\end{equation*}

\noindent
{\bf{Example~2.}} Consider $X\sim \text{Exp} (\lambda ),~\lambda>0$. Then,
\begin{equation}
	\mathbb{E}[R_n]
	= \frac{1}{\lambda}  \sum_{k=1}^{n-1} \frac{1}{ k}  = O \left( \frac{1}{\lambda}\log(n) \right).
\end{equation}

\noindent
{\bf{Example~3.}} Let $X$ be $\gamma^2$-sub-Gaussian\footnote{A random variable $X$ is $\gamma^2$-sub-Gaussian if $\mathbb{E}[e^{\lambda(X-\mathbb{E}[X])}] \le e^{\lambda \gamma^2}$ for all $\lambda \in \mathbb{R}$. }.  Then \cite{boucheron2013concentration},
\begin{equation*}
	\mathbb{E}[R_n] \le 2 \sqrt{ 2 \gamma^2 \log(n) }.
\end{equation*}

\appendices
\section{Proof of Theorem~\ref{prop:sigto0}}
\label{app:LowNoise}
Before proceeding with the proof of Theorem~\ref{prop:sigto0}, we first present two ancillary results. 

\begin{lemma}\label{lem:boundPDFw}
Let $\Xm$ consist of $n$ i.i.d. random variables generated according to $X$.  Let $X'$ be an independent copy of $X$ and assume that
\begin{equation*}
 f_{X-X^\prime}(x)< \infty,~\forall x \in \mathbb{R}. 
\end{equation*} 
Then, the following holds
\begin{align*}
	& f_{W_i}(u) < \infty,~\forall u,\in\mathbb{R}_+, 1\le i \le n-1,\\
	& f_{W_i,W_j}(u,v) <\infty,~\forall(u,v)\in\mathbb{R}_+^2, \, 1\le i<j \le n-1,
\end{align*}
where $W_i = X_{i+1:n}  - X_{i:n},~i\in[1:n-1]$.
\end{lemma}

\begin{IEEEproof}
The proof is provided in Appendix~\ref{app:ProofLemmaboundPDFw}.
\end{IEEEproof}

\begin{lemma} 
\label{lemma:NA}
Let  $\Vm\sim\Nc(\mathbf{0}_{n-1}, \tilde{K})$ where $\tilde{K}$ is defined in~\eqref{eq:KTilde}.  Then, for any subset $\Ic \subseteq [1:n-1]$, 
\begin{equation}
\Pr \left( \bigcap_{i\in \Ic}  \{ V_i \le t_i \}  \right) \le \prod_{i\in \Ic}   \Pr \left(  \{ V_i \le t_i \}  \right).  \label{eq:Negative_association_inequality}
\end{equation} 
\end{lemma}
\begin{IEEEproof}
The bound in \eqref{eq:Negative_association_inequality} holds if the random  vector $\Vm$ consists of {\em negatively associated} random variables~\cite{joag1983negative}. 
Observe that the Gaussian random vector $\Vm\sim\Nc(\mathbf{0}_{n-1}, \tilde{K})$ consists of either negatively correlated or independent random variables (see the structure of $\tilde{K}$ in~\eqref{eq:KTilde}).  As was shown in  \cite{joag1983negative}, this implies that the random variables in ${\Vm}$ are negatively associated. 
This concludes the proof of Lemma~\ref{lemma:NA}.
\end{IEEEproof}

\smallskip

We now leverage the two lemmas above to prove Theorem~\ref{prop:sigto0}. 
From Corollary~\ref{prop:Peiso} we have that
\begin{align}\label{eq:sig0pf1}
	P_e 
	& = 1 - \mathbb{E}\left[ \Pr \left(\Vm\ge - \frac{T_{\tau}\Xm}{\sigma} \;\middle | \;   \Xm \right) \;\middle | \;    {\bf X} \in \mathcal{H}_\tau  \right],
\end{align}
where  $\Vm\sim\Nc(\mathbf{0}_{n-1}, \tilde{K})$.
The expression in~\eqref{eq:sig0pf1} can be equivalently written as
\begin{align}\label{eq:sig0pf2}
	P_e 
	& = \mathbb{E} \left[ 1 \!-\! \Pr \!\left( \bigcap_{i=1}^{n-1} \left\{ V_i \ge \frac{X_{\tau_i} \!-\! X_{\tau_{i+1}} }{\sigma} \right\} \;\middle | \;   \Xm \right) \; \middle | \; \Xm\in\Hc_\tau \right] \nonumber \\
	& = \mathbb{E} \left[ \Pr\left( \bigcup_{i=1}^{n-1} \left\{ V_i < \frac{X_{\tau_i} \!-\! X_{\tau_{i+1}} }{\sigma} \right\} \;\middle | \;   \Xm \right) \; \middle | \; \Xm\in\Hc_\tau \right] \nonumber \\
	& \overset{\rm (a)}{=} \Pr\left( \bigcup_{i=1}^{n-1} \left\{ V_i < \frac{X_{\tau_i} \!-\! X_{\tau_{i+1}} }{\sigma} \right\} \;\middle | \;   \Xm\in\Hc_\tau \right) \nonumber \\
	& \overset{\rm (b)}{=}  \sum_{k=1}^{n-1} \left( (-1)^{k-1} \! \sum_{\substack{\Ic \subseteq  [1:n-1] \\ | \Ic |=k}}  \Pr\left( \Ac_{\Ic} \right)  \right) ,
\end{align}
where $\rm{(a)}$ is due to the law of total expectation, and $\rm{(b)}$ follows from the inclusion-exclusion principle where $\Ac_{\Ic} = \cap_{i\in \Ic}\Ac_i$ with $\Ac_i = \{   V_i < \sigma^{-1} (X_{\tau_i} - X_{\tau_{i+1}}) \mid \Xm\in\Hc_\tau \}$.

From the expression in~\eqref{eq:sig0pf2} it follows that
\begin{align}
\label{eq:GenLimPe}
	\lim_{\sigma\to0} \frac{P_e}{\sigma}
	& =  \sum_{k=1}^{n-1} \left( (-1)^{k-1} \sum_{\substack{\Ic \subseteq  [1:n-1] \\ | \Ic |=k}} \lim_{\sigma\to0}  \frac{1}{\sigma} \Pr\left( \Ac_{\Ic} \right)  \right).
\end{align}
In what follows, we therefore analyze $\Pr\left( \Ac_{\Ic} \right)$. We have that  
\begin{align}\label{eq:sig0pf5}
	\Pr\left( \Ac_{\Ic} \right)   
	& = \mathbb{E} \left[   \Pr\!\left( \bigcap_{i\in \Ic} \left\{ V_i < \frac{X_{\tau_{i}} \!-\! X_{\tau_{i+1}}}{\sigma} \right\}\! \;\middle | \;  \! \Xm \right)  \! \; \middle | \;\! \Xm\in\Hc_\tau \right] \nonumber \\
	& = \mathbb{E} \left[   \Pr\!\left( \bigcap_{i\in \Ic} \left\{ V_i < \frac{X_{i:n} \!-\! X_{i+1:n}}{\sigma} \right\} \;\middle | \;   \Xm \right)  \right] \nonumber \\
	& = \mathbb{E} \left[   \Pr\!\left( \bigcap_{i\in \Ic} \left\{ V_i < \frac{- W_i}{\sigma} \right\}  \;\middle | \; \Wm_{\Ic} \right)    \right],
\end{align}
where $\Wm_{\Ic}$ is a $k$-dimensional random vector with entries $W_i =X_{i+1:n} -X_{i:n}$ for $i \in \Ic$.

We next consider two separate cases.

\noindent $\bullet$ {\bf{Case 1:}} $k = 1$. Let $\Ic = \{i\}$; then, we can write~\eqref{eq:sig0pf5} as
\begin{align*}
\Pr\left( \Ac_{\Ic} \right) &= \mathbb{E} \left[   \Pr \left(   V_i < \frac{- W_i}{\sigma}   \;\middle | \; W_i \right)    \right]
\\ & =  \mathbb{E} \left[   Q\left( \frac{W_i}{\sqrt{2}\sigma} \right)  \right]
\\ &=   \int_{0}^\infty    Q\left( \frac{w}{\sqrt{2}\sigma} \right)  f_{W_i}\left( w \right) {\rm d} w \\
		  &=  \int_{0}^\infty    Q\left( u \right)  f_{W_i}\left( \sqrt{2}\sigma u \right) \sqrt{2}\sigma \ {\rm d} u ,
\end{align*}
where the last equality follows by applying the change of variable. 
Thus, we have that
\begin{align}\label{eq:sig0k1}
	 \lim_{\sigma\to0} \frac{1}{\sigma} \Pr\left( \Ac_{\Ic} \right) 
	& = \sqrt{2} \int_{0}^\infty  \lim_{\sigma\to0}  Q\left( u \right)  f_{W_i}\left( \sqrt{2}\sigma u \right)  {\rm d} u \nonumber \\
	&= \sqrt{2}  f_{W_i}\left( 0^+ \right)  \int_{0}^\infty   Q\left( u \right)   {\rm d} u \nonumber \\
	& =  \frac{f_{W_i}\left( 0^+ \right)}{\sqrt{\pi}},
\end{align}
where the first equality follows from the dominated convergence theorem, which is verifiable since for any $\sigma$, $Q(u)f_{W_i}(\sqrt{2}\sigma u) \le Q(u) \max_{t} f_{W_i}(t) <\infty$  where the fact that $f_{W_i}(t)<\infty,~\forall t \in \mathbb{R}_+$ is shown in Lemma~\ref{lem:boundPDFw}. \hfill $\square$

\noindent $\bullet$ {\bf{Case 2:}} $k \geq 2$. 
By using the bound in Lemma~\ref{lemma:NA}, we obtain
\begin{align}
	\Pr\left( \Ac_{\Ic} \right)   
	& \le \mathbb{E} \left[  \prod_{i\in \Ic} \Pr\!\left( V_i < \frac{- W_i}{\sigma} \;\middle | \; \Wm_{\Ic} \right)    \right] \nonumber \\
	& = \mathbb{E} \left[  \prod_{i\in \Ic} Q\!\left(\frac{ W_i}{\sqrt{2}\sigma} \right)    \right] \nonumber \\
	&\le \mathbb{E} \left[  \prod_{\substack{i\in \Jc \\ \Jc \subset \Ic, |\Jc|=2}} Q\!\left(\frac{ W_i}{\sqrt{2}\sigma} \right)    \right]. \label{eq:upQ1}
\end{align}
By letting $\Jc = \{s,t\} \subset \Ic$ in~\eqref{eq:upQ1}, we obtain that
\begin{align}
	& \lim_{\sigma \to 0}\frac{\Pr\left( \Ac_{\Ic} \right) }{ \sigma}  \notag\\ 
	& \le \lim_{\sigma \to 0}   \sigma \!\! \int_0^\infty \!\!\! \int_0^\infty \!\!Q \! \left(\frac{w}{\sqrt{2}} \right) \! Q\! \left(\frac{z}{\sqrt{2}} \right) \! f_{W_{s},W_{t}}(\sigma w,\sigma z) \ {\rm d}w \ {\rm d}z \nonumber
\\& \le \lim_{\sigma\to0} \sigma \frac{f_{W_{s},W_{t}}(0^+,0^+)}{\pi} = 0,
\label{eq:i2limit}
\end{align}
where the equality follows from Lemma~\ref{lem:boundPDFw}, and the second inequality is due to the fact that
\begin{align}\label{eq:limint}
	& \lim_{\sigma\to0} \int_0^\infty \!\!\! \int_0^\infty \!\!Q \! \left(\frac{w}{\sqrt{2}} \right) \! Q\! \left(\frac{z}{\sqrt{2}} \right) \! f_{W_{s},W_{t}}(\sigma w,\sigma z) \ {\rm d}w \ {\rm d}z \nonumber \\
	& \stackrel{{\rm{(a)}}}{=} \int_0^\infty \!\!\! \int_0^\infty \!\!Q \! \left(\frac{w}{\sqrt{2}} \right) \! Q\! \left(\frac{z}{\sqrt{2}} \right) \! f_{W_{s},W_{t}}(0^+,0^+) \ {\rm d}w \ {\rm d}z \nonumber \\
	&= \frac{f_{W_{s},W_{t}}(0^+,0^+)}{\pi} < \infty,
\end{align}
where $\rm{(a)}$ follows from the dominated convergence theorem, which is verifiable by means of Lemma~\ref{lem:boundPDFw}. \hfill $\square$

By using the limits in~\eqref{eq:sig0k1} and~\eqref{eq:i2limit} inside~\eqref{eq:GenLimPe}, we obtain
\begin{align}
	\lim_{\sigma\to0} \frac{P_e}{\sigma}
	& \overset{\rm (a)}{=}  \sum_{\substack{\Ic \subseteq [1:n-1] \\ | \Ic |=1}}   \lim_{\sigma\to0} \frac{1}{\sigma}  \Pr\left( \Ac_{\Ic} \right)  \nonumber \\
	& \overset{\rm (b)}{=} \sum_{i=1}^{n-1} \frac{f_{W_i}\left( 0^+ \right)}{\sqrt{\pi}},
\end{align}
where $\rm (a)$ follows from~\eqref{eq:i2limit}, and $\rm (b)$ follows from~\eqref{eq:sig0k1}.
This concludes the proof of Theorem~\ref{prop:sigto0}.

\section{Proof of Theorem~\ref{thm:sigtoInf}}
\label{app:HighNoise}
We start by noting that, in view of the limit in~\eqref{eq:Limit_sigma_infinity}, we have that $\lim_{\sigma \to \infty} P_c=\frac{1}{n!}$.  We now consider the following limit,
\begin{equation}\label{eq:siginfrate}
	\lim_{\sigma\to\infty} \frac{P_c - \frac{1}{n!}}{\frac{1}{\sigma}}.
\end{equation}
Instead of working with $\sigma$, we parameterize the problem in terms of  $\sigma=\frac{1}{\kappa}$. 
Then, \eqref{eq:siginfrate} can be equivalently expressed as
\begin{align}\label{eq:derkappa}
	\lim_{\sigma\to\infty}  \frac{P_c - \frac{1}{n!}}{\frac{1}{\sigma}}
	 = \lim_{\kappa\to0}  \frac{P_c - \frac{1}{n!}}{\kappa}
	 = \lim_{\kappa\to0} \frac{\partial P_c }{\partial \kappa},
\end{align}
where the last equality can be argued by using the definition of the derivative or the  L'H\^opital's rule.

From Corollary~\ref{prop:Peiso}, the probability of correctness is given by
\begin{align}
	P_c = 1-P_e
& = \mathbb{E}\left[ Q_{\sigma^2\tilde{K}} \! \left( -  T_\tau \Xm \right)  \;\middle | \;    {\bf X} \in \mathcal{H}_\tau  \right] \nonumber
\\
	& = \mathbb{E}\left[ \Pr \left( \frac{1}{\kappa} \Vm \ge - \Wm  \;\middle | \;  \Wm \right)   \right] \nonumber\\
	 &= \mathbb{E}\left[ \int_{\vv\in\mathbb{R}^{n-1}}  \! 1_{ \left\{ \vv \ge - \kappa \Wm \right\}} f_{{\Vm}}(\vv)   {\rm d}\vv    \right], \label{eq:Double_integration_expression}
\end{align}
where we let  $\Vm \sim\Nc(\mathbf{0}_{n-1},  \tilde{K})$ and $\Wm = T_\tau \Xm \mid \Xm\in\Hc_\tau$, and we used the exchangeablity of $\Xm$.

Using the expression in~\eqref{eq:Double_integration_expression}, the derivative of $P_c$ with respect to $\kappa$ is now given by 
\begin{align}\label{eq:derivPc1}
	 \frac{\partial P_c }{\partial \kappa}
	& \overset{\rm (a)}{=} \mathbb{E}\left[ \int_{\vv\in\mathbb{R}^{n-1}}  \! \frac{\partial  }{\partial \kappa} 1_{ \left\{ \vv \ge -\kappa{\Wm} \right\}} f_{{\Vm}}(\vv)   {\rm d}\vv   \right] \nonumber \\
	& \overset{\rm (b)}{=}  \mathbb{E}\left[ \int_{\vv\in\mathbb{R}^{n-1}}  \! \bigtriangleup(\kappa,\vv,\Wm) f_{{\Vm}}(\vv)   {\rm d}\vv    \right],
\end{align}
where in $\rm (a)$ we used the Leibniz's integral rule, 
and $\rm (b)$ follows since
\begin{align}\label{eq:delfc}
	 \frac{\partial  }{\partial \kappa} 1_{ \left\{ \vv \ge -\kappa{\Wm} \right\}} 
	& = \frac{\partial  }{\partial \kappa} \prod_{i=1}^{n-1} 1_{ \left\{ v_i \ge -\kappa W_{i}  \right\}} \nonumber \\
	& {\overset{\rm (b1)}{=}} \frac{\partial  }{\partial \kappa} \prod_{i=1}^{n-1} 1_{  \left\{ \frac{ - v_i}{W_{i}} \le \kappa   \right\}} \nonumber \\
	& {\overset{\rm (b2)}{=}}   \sum_{i=1}^{n-1} \delta \left(\kappa + \frac{ v_i}{W_{i}} \right) \prod_{\substack{j=1 \\ j\neq i }}^{n-1} 1_{  \left\{ \frac{ - v_j}{W_{j}} \le \kappa   \right\}} \nonumber \\
	& {\overset{\rm (b3)}{=}}  \sum_{i=1}^{n-1} W_{i} \delta \left( \kappa W_{i} +  v_i \right) \prod_{\substack{j=1 \\ j\neq i }}^{n-1} 1_{  \left\{ \frac{ - v_j}{W_{j}} \le \kappa   \right\}} \nonumber \\
	& \triangleq \bigtriangleup(\kappa,\vv,\Wm),
\end{align}
where the labeled equalities follow from: $\rm (b1)$ since each entry $W_i, i \in [1:n-1]$ of $\Wm$ is positive (we ignore the case when $(T_\tau\Xm)_i=0$, and in fact $f_{W_i}(0)=0$ for an i.i.d. $\Xm$); $\rm (b2)$ the product rule and the fact that $\frac{\partial}{\partial x} 1_{\{ 0 \le x+t \}} = \delta ( x+t )$, where $\delta(x)$ is the Dirac delta function; $\rm (b3)$ the scaling property of the Dirac delta function.

We now consider the integral inside the expectation in~\eqref{eq:derivPc1}. By using the sifting property of the Dirac delta function, the integral becomes
\begin{align}\label{eq:derivPc2}
	& \int_{\vv\in\mathbb{R}^{n-1}}  \! \bigtriangleup(\kappa,\vv,\Wm) f_{{\Vm}}(\vv)   {\rm d}\vv \nonumber \\
	& \!=\!  \sum_{i=1}^{n-1} W_{i} \! \int_{\uv\in\mathbb{R}^{n-2}}  \! \prod_{\substack{j=1 \\ j \neq i}}^{n-2} \!1_{  \left\{ \frac{ - u_j}{W_j} \le \kappa   \right\}}  f_{{\Vm}_{\setminus i}, V_i}(\uv,- \kappa W_i )    {\rm d}\uv  ,
\end{align}
where ${\Vm}_{\setminus i}$ is obtained by retaining all the entries of $\Vm$ except the $i$-th one.
We next substitute~\eqref{eq:derivPc2} inside~\eqref{eq:derivPc1} and we compute the limit in~\eqref{eq:derkappa}. We obtain
\begin{align}
\label{eq:LimDerPc}
	\lim_{\kappa\to0} \frac{\partial P_c}{\partial \kappa}
	& \overset{\rm (a)}{=}  \mathbb{E}  \! \left[ \sum_{i=1}^{n-1} W_{i}  \int_{\uv\in\mathbb{R}^{n-2}}  \!\!\prod_{\substack{j=1 \\ j \neq i}}^{n-2} \! 1_{  \left\{  u_j \ge 0   \right\}} f_{{\Vm}_{\setminus i}, V_i}(\uv,0 )   \  {\rm d}\uv   \right] \nonumber \\
	& =   \sum_{i=1}^{n-1}\mathbb{E}  \left[ W_{i}    \right]  \int_{\uv\in\mathbb{R}^{n-2}_+}    f_{{\Vm}_{\setminus i}, V_i}(\uv,0 ) \  {\rm d}\uv \nonumber \\
	& \stackrel{\rm (b)}{=}  \!  \sum_{i=1}^{n-1} \!  \mathbb{E} \! \left[  W_i \right]  \! \Pr(  {\Vm}_{\setminus i} \!\ge \! \mathbf{0}_{n-2}  \mid V_i=0) f_{V_i}(0),
\end{align}
where the labeled equalities follows from: $\rm (a)$ 
using the dominated convergence theorem, which is verified since $W_i \prod_{\substack{j=1 \\ j \neq i}}^{n-2} 1_{  \left\{  u_j \ge 0   \right\}} f_{{\Vm}_{\setminus i}, V_i}(\uv,0 ) \leq W_i  f_{{\Vm}_{\setminus i}, V_i}(\uv,0 )$ where $W_{i} $ is assumed to be absolutely integrable; 
and $\rm (b)$ using the following,
\begin{align*}
\int_{\uv\in\mathbb{R}^{n-2}_+}    f_{{\Vm}_{\setminus i}, V_i}(\uv,0 ) \  {\rm d}\uv &= \int_{\uv\in\mathbb{R}_+^{n-2}}  \! f_{{\Vm}_{\setminus i} | V_i }(\uv\mid 0) f_{V_i}(0)   {\rm d}\uv
\\& = \Pr(  {\Vm}_{\setminus i} \ge  \mathbf{0}_{n-2}  | V_i=0) f_{V_i}(0).
\end{align*}
To finalize the proof, it remains to compute  $\Pr(  {\Vm}_{\setminus i} \ge  \mathbf{0}_{n-2}  | V_i=0) =\alpha_i$. This can be done as follows,
\begin{align}
	& \alpha_i 
	= \Pr(  {\Vm}_{\setminus i} \ge  \mathbf{0}_{n-2} \mid V_i=0) \nonumber\\
	& \stackrel{\rm (a)}{=}  \Pr \!\left( \bigcap_{{j=1 , j\neq i }}^{n-1} \{ Z_{j+1}-Z_j \ge 0 \}  \; \middle | \; Z_{i+1}-Z_i=0 \right)  \nonumber\\
	& =\Pr \!\left(  \{ Z_1\le \cdots \le Z_{i} \} \cap  \{ Z_{i+1}\le \cdots \le Z_{n} \} \mid Z_{i+1} = Z_i \right )  \nonumber\\
	& \stackrel{\rm (b)}{= }  \mathbb{E} \! \left[ \Pr \!\left(  \{  \cdots \le Z_{i} \} \cap  \{ Z_{i+1}\le \cdots\} | Z_{i+1} ,Z_i \right ) | Z_{i+1} = Z_i \right]  \nonumber\\
	& = \int_{-\infty}^{\infty} \Pr \!\left(  \{  \cdots \le t \} \cap  \{ t \le \cdots  \}  \right ) f_{Z_i,Z_{i+1}\mid Z_i=Z_{i+1}} (t,t) \ {\rm d} t \nonumber\\
	& \stackrel{\rm (c)}{= } \int_{-\infty}^{\infty} \Pr \!\left(  \{ Z_1\le \cdots \le t \} \cap  \{ t \le \cdots \le Z_{n} \}  \right ) f_{\frac{1}{\sqrt{2}}{Z}} (t) \ {\rm d} t \nonumber\\
	& =  \Pr \!\left(   Z_1\le \cdots \le Z_{i-1} \le \frac{1}{\sqrt{2}}{Z}_i \le Z_{i+2} \le \cdots \le Z_{n}   \right ) \nonumber  \\
	&  \stackrel{\rm (d)}{= }    \Pr \!\left( A_i {\bf Z}  \in  \Hc_{[1:n-1]}   \right ) \nonumber  \\
	& =    \Pr \!\left(  {\bf Z}  \in A_i^{-1}  \Hc_{[1:n-1]}   \right ) \nonumber \\
	& \overset{\rm (e)}{=}  \frac{ {\rm Vol}\left( \Bc (\mathbf{0}_{n-1},1) \cap A_i^{-1}  \Hc_{[1:n-1]}  \right )}{{ \rm Vol} \left (\Bc(\mathbf{0}_{n-1},1) \right)}\nonumber \\
	& \overset{\rm (f)}{=} \left | \det\left( A_i^{-1} \right)  \right | \frac{ {\rm Vol}\left( A_i \Bc (\mathbf{0}_{n-1},1) \cap  \Hc_{[1:n-1]} \right )}{{ \rm Vol} \left (\Bc(\mathbf{0}_{n-1},1) \right)} \nonumber \\
	& \overset{\rm (g)}{=} \sqrt{2}  \frac{ {\rm Vol}\left( \Ec (\mathbf{0}_{n-1},i) \cap  \Hc_{[1:n-1]} \right )}{{ \rm Vol} \left (\Bc(\mathbf{0}_{n-1},1) \right)}, \label{eq:PrVol}
 \end{align} 
where the labeled equalities follow from:
$\rm (a)$ the definition of  ${\Vm}$ and writing it in terms of standard normal; 
$\rm (b)$ the law of total expectation, where we abbreviate  $\{  \cdots \le Z_{i} \} \cap  \{ Z_{i+1}\le \cdots\} \triangleq  \{ Z_1\le \cdots \le Z_{i} \} \cap  \{ Z_{i+1}\le \cdots \le Z_{n} \} $;
$\rm (c)$ using the fact that
\begin{align*}
	f_{Z_i,Z_{i+1}\mid Z_i=Z_{i+1}} (t,t)
	& = \frac{f_{Z_i,Z_{i+1}} (t,t) }{ \int_{-\infty}^{\infty}f_{Z_i,Z_{i+1}} (z,z) \ {\rm d} z}  \nonumber \\
	& = f_{\frac{1}{\sqrt{2}}{Z}} (t);
\end{align*}
$\rm (d)$ letting ${\bf{Z}} \sim \mathcal{N}(\mathbf{0}_{n-1},I_{n-1})$, defining a diagonal matrix $A_i \in \mathbb{R}^{(n-1) \times (n-1)}$ with the $i$-th element equal to $\frac{1}{\sqrt{2}}$ and the others equal to one, and recalling  that from~\eqref{eq:HR} we have $\Hc_{[1:n-1]} = \{\xv \in \mathbb{R}^{n-1}: x_1\le \cdots \le x_{n-1}\}$; 
$\rm (e)$ using the $(n-1)$-dimensional volume expression for the probability of a standard normal vector~\cite{jeong20};
$\rm (f)$ the fact that ${\rm Vol} (A\Sc ) = | \det(A) | {\rm Vol} (\Sc )$ for any invertible matrix $A$ and any set $\Sc$;
$\rm (g)$ letting $\Ec ( \mathbf{0}_{n-1},i)$ be the $(n-1)$-dimensional ellipsoid centered at the origin with unit radii along standard axes except a $\frac{1}{\sqrt{2}}$ radius along the $i$-th axis.

Substituting~\eqref{eq:PrVol} into~\eqref{eq:LimDerPc}, and noting that  $ f_{V_i}(0)=\frac{1}{2 \sqrt{\pi}}$ for all $i \in [1:n-1]$, we obtain
\begin{align}
	\lim_{\sigma \to \infty} \frac{P_e(\infty) - P_e(\sigma)}{\frac{1}{\sigma}}
	& = \frac{1}{\sqrt{2 \pi}} \sum_{i=1}^{n-1} \alpha_i   \mathbb{E}  \left[  W_i \right],
\end{align}
where, for all $i \in [1:n-1]$, we have
\begin{align}
	\alpha_i
	& =  \frac{ {\rm Vol}\left( \Ec (\mathbf{0}_{n-1},i) \cap  \Hc_{[1:n-1]} \right )}{{ \rm Vol} \left (\Bc(\mathbf{0}_{n-1},1) \right)}.
\end{align}
This concludes the proof of Theorem~\ref{thm:sigtoInf}.

\section{Proof of Lemma~\ref{lem:boundPDFw}}
\label{app:ProofLemmaboundPDFw}
We start by noting that the joint density function $f_{W_i,W_j}(u,v),~1\le i<j\le n-1$ is given by~\cite{spacingPyke}
\begin{align*}
	& f_{W_i,W_j}(u,v) \nonumber\\
	& = n! \!\int_{-\infty}^{\infty} \int_{x+u}^\infty \! \frac{F(y)^{i-2}}{(i-2)!} \frac{(F(x)\!-\!F(x+u))^{j-i-2}}{(j-i-2)!} \nonumber\\
	&~~~ \times \!\frac{(1\!-\!F(y\!+\!v))^{n-j}}{(n-j)!} f(x) f(x\!+\!u) f(y) f(y\!+\!v) \ {\rm d} y \ {\rm d} x,
\end{align*}
where $F(\cdot)$ is the cumulative distribution function of $X$ and $f(\cdot)$ is the probability density function of $X$. 

By using the upper bounds of $F(x)\le1$ and $1-F(x) \le 1$, we obtain
\begin{align*}
	& f_{W_i,W_j}(u,v) \nonumber\\
	& \le n!  \frac{\int_{-\infty}^{\infty} \int_{x+u}^\infty f(x) f(x+u) f(y) f(y+v)  \ {\rm d} y \ {\rm d} x}{(i-2)!(j-i-2)!(n-j)!} \nonumber \\
	& \le n!  \frac{\int_{-\infty}^{\infty} \int_{-\infty}^\infty f(x) f(x+u) f(y) f(y+v)  \ {\rm d} y\ {\rm d} x }{(i-2)!(j-i-2)!(n-j)!} \nonumber \\
	& = n!  \frac{\int_{-\infty}^{\infty} f(x) f(x+u)\ {\rm d} x \int_{-\infty}^\infty  f(y) f(y+v) \ {\rm d} y  }{(i-2)!(j-i-2)!(n-j)!} \nonumber \\
	& =  \frac{n!  f_{X-X^\prime}(u) f_{X-X^\prime}(v) }{(i-2)!(j-i-2)!(n-j)!} \\
	& < \infty , 
\end{align*}
where the second inequality follows since the integrand is always positive, and the last inequality is due to the assumption that $f_{X-X^\prime}(x)<\infty,~\forall x$.
This shows that the joint density is bounded everywhere.

For the marginal density, we obtain
\begin{align*}
	f_{W_i}(u) 
	& = \int_{-\infty}^{\infty} \! f_{W_i,W_j}(u,v) \ {\rm d}v \nonumber \\
	& \le \int_{-\infty}^{\infty} \! \frac{ n!   f_{X-X^\prime}(u) f_{X-X^\prime}(v) }{(i-2)!(j-i-2)!(n-j)!} \ {\rm d}v \nonumber \\
	& =  \frac{n!  f_{X-X^\prime}(u)  }{(i-2)!(j-i-2)!(n-j)!} \nonumber \\
	& < \infty,
\end{align*}
where the inequality follows from the fact that we have shown above that
\begin{align*}
f_{W_i,W_j}(u,v) \leq \frac{n!  f_{X-X^\prime}(u) f_{X-X^\prime}(v) }{(i-2)!(j-i-2)!(n-j)!}.
\end{align*}
This concludes the proof of Lemma~\ref{lem:boundPDFw}.

\section{Examples for the Low Noise Regime}
\subsection{Uniform Distribution}
\label{app:SlopeUnif}
For  $X\sim \text{Unif}(a,b),~0\le a<b<\infty$, using the formula in~\eqref{eq:fWi}, we have that
\begin{align*}
	f_{W_i}(0^+) 
	& = \frac{n! \int_{a}^{b} \left(\frac{x-a}{b-a}\right)^{i-1} \left(  \frac{b-x}{b-a} \right)^{n-i-1} ( b-a)^{-2} \ {\rm d} x}{(i-1)! (n-i-1)!}  \nonumber \\
	& = \frac{n! \int_{a}^{b} \left(x-a\right)^{i-1} \left( b-x \right)^{n-i-1}  \ {\rm d} x}{(b-a)^n (i-1)! (n-i-1)!}  \nonumber \\
	& = \frac{n!  }{(b-a)^n (i-1)! (n-i-1)! } \frac{  \Gamma(n-i) \Gamma(i) }{(b-a)^{1-n}\Gamma(n) }  \nonumber \\
	& = \frac{n   }{ b-a  }, 
\end{align*}
where $\Gamma (\cdot)$ is the gamma function.
Hence,~\eqref{eq:lims0} becomes
\begin{align*}
	\lim_{\sigma\to0} \frac{P_e}{\sigma}
	 =   \sum_{i=1}^{n-1} \frac{f_{W_i}(0^+) }{\sqrt{\pi}}  
	 =  \frac{n (n-1)}{(b-a)\sqrt{\pi}}.
\end{align*}

\subsection{Exponential Distribution}
\label{app:SlopeExp}
For the case of $X\sim \text{Exp}(\lambda)$, using the formula in~\eqref{eq:fWi}, we have that
\begin{align*}
	f_{W_i}(0^+) 
	& = \frac{\lambda^2 n! \int_{0}^{\infty} (1-e^{-\lambda x})^{i-1} e^{-(n-i-1)\lambda x}  e^{-2\lambda x} \ {\rm d} x}{(i-1)! (n-i-1)!}  \nonumber\\
	& = \frac{\lambda^2 n! \int_{0}^{\infty} (1-e^{-\lambda x})^{i-1} e^{-(n-i+1)\lambda x}   \ {\rm d} x}{(i-1)! (n-i-1)!}\nonumber \\
	& = \frac{ \lambda^2 n!  }{  (i-1)! (n-i-1)! } \frac{\Gamma(n-i+1) \Gamma(i)}{ \lambda \Gamma(1+n)} \nonumber \\
	& = \frac{ \lambda^2 n!  }{  (i-1)! (n-i-1)! } \frac{(n-i)! (i-1)! }{ \lambda n!} \nonumber \\
	& = \lambda (n-i),
\end{align*}
where $\Gamma (\cdot)$ is the gamma function.
Hence, \eqref{eq:lims0} becomes
\begin{align*}
	\lim_{\sigma\to0} \frac{P_e}{\sigma}
	 =   \sum_{i=1}^{n-1} \frac{f_{W_i}(0^+)}{\sqrt{\pi}} 
	 = \frac{\lambda (n-1)n}{2\sqrt{\pi}}.
\end{align*}

\section{Examples for the High Noise Regime}
\label{app:ExamHighNoise}
The key to the proof is to use the following expressions from~\cite{orderstat},
\begin{align*}
	& \mathbb{E}[X_{1:n}] 
	= n \int_{-\infty}^{\infty} x (1-F(x))^{n-1} f(x) \ {\rm d} x  ,\\
	& \mathbb{E}[X_{n:n}] 
	= n \int_{-\infty}^{\infty} x F(x)^{n-1} f(x) \ {\rm d} x .
\end{align*}
First, consider $X_i \sim \text{Unif}(a,b),~0\le a< b <\infty$. Then, 
\begin{align*}
	\mathbb{E}[X_{1:n}] 
	 = \frac{b+an}{n+1} \ \text{ and } \
	\mathbb{E}[X_{n:n}] 
	 = \frac{a+bn}{n+1},
\end{align*}
and hence, we obtain
\begin{align*}
	\mathbb{E}[R_n]
	= \frac{(b-a)(n-1)}{n+1}.
\end{align*}
Next, let $X_i\sim \text{Exp}(\lambda)$. Then,
\begin{align*}
	\mathbb{E}[X_{1:n}] 
	 = \frac{1}{\lambda n} \ \text{ and } \
	\mathbb{E}[X_{n:n}] 
	 = \sum_{k=1}^n \frac{1}{\lambda k},
\end{align*}
and hence, we obtain
\begin{align*}
	\mathbb{E}[R_n]
	= \sum_{k=1}^{n-1} \frac{1}{\lambda k}.
\end{align*}

\bibliographystyle{IEEEtran}
\bibliography{HypoTestNetwork}

\begin{thebibliography}{10}
\providecommand{\url}[1]{#1}
\csname url@samestyle\endcsname
\providecommand{\newblock}{\relax}
\providecommand{\bibinfo}[2]{#2}
\providecommand{\BIBentrySTDinterwordspacing}{\spaceskip=0pt\relax}
\providecommand{\BIBentryALTinterwordstretchfactor}{4}
\providecommand{\BIBentryALTinterwordspacing}{\spaceskip=\fontdimen2\font plus
\BIBentryALTinterwordstretchfactor\fontdimen3\font minus
  \fontdimen4\font\relax}
\providecommand{\BIBforeignlanguage}[2]{{%
\expandafter\ifx\csname l@#1\endcsname\relax
\typeout{** WARNING: IEEEtran.bst: No hyphenation pattern has been}%
\typeout{** loaded for the language `#1'. Using the pattern for}%
\typeout{** the default language instead.}%
\else
\language=\csname l@#1\endcsname
\fi
#2}}
\providecommand{\BIBdecl}{\relax}
\BIBdecl

\bibitem{dwork2008differential}
C.~Dwork, ``Differential privacy: {A} survey of results,'' in
  \emph{International conference on theory and applications of models of
  computation}.\hskip 1em plus 0.5em minus 0.4em\relax Springer, 2008, pp.
  1--19.

\bibitem{jeong20}
M.~{Jeong}, A.~{Dytso}, M.~{Cardone}, and H.~V. {Poor}, ``Recovering structure
  of noisy data through hypothesis testing,'' in \emph{Proceedings of the 2020
  IEEE International Symposium on Information Theory (ISIT)}, June 2020, pp.
  1307--1312.

\bibitem{ourJSAIT}
------, ``Recovering data permutations from noisy observations: The linear
  regime,'' \emph{IEEE Journal on Selected Areas in Information Theory},
  vol.~1, no.~3, pp. 854--869, 2020.

\bibitem{searle1973prediction}
S.~R. Searle \emph{et~al.}, ``Prediction, mixed models, and variance
  components,'' 1973.

\bibitem{portnoy1982maximizing}
S.~Portnoy, ``Maximizing the probability of correctly ordering random variables
  using linear predictors,'' \emph{Journal of Multivariate Analysis}, vol.~12,
  no.~2, pp. 256--269, 1982.

\bibitem{nomakuchi1988lemma}
K.~Nomakuchi and T.~Sakata, ``Characterizations of the forms of covariance
  matrix of an elliptically contoured distribution,'' \emph{Sankhy{\=a}: The
  Indian Journal of Statistics, Series A}, pp. 205--210, 1988.

\bibitem{nomakuchi1988}
------, ``Characterization of conditional covariance and unified theory in the
  problem of ordering random variables,'' \emph{Annals of the Institute of
  Statistical Mathematics}, vol.~40, no.~1, pp. 93--99, 1988.

\bibitem{Collier}
O.~Collier and A.~S. Dalalyan, ``Minimax rates in permutation estimation for
  feature matching,'' \emph{The Journal of Machine Learning Research}, vol.~17,
  no.~6, pp. 1 --31, January 2016.

\bibitem{Pananjady2018}
A.~{Pananjady}, M.~J. {Wainwright}, and T.~A. {Courtade}, ``Linear regression
  with shuffled data: Statistical and computational limits of permutation
  recovery,'' \emph{IEEE Transactions on Information Theory}, vol.~64, no.~5,
  pp. 3286--3300, May 2018.

\bibitem{Pananjady2017}
------, ``Denoising linear models with permuted data,'' in \emph{Proceedings of
  the 2017 IEEE International Symposium on Information Theory (ISIT)}, June
  2017, pp. 446--450.

\bibitem{rigollet2019uncoupled}
P.~Rigollet and J.~Weed, ``Uncoupled isotonic regression via minimum
  {W}asserstein deconvolution,'' \emph{Information and Inference: A Journal of
  the IMA}, vol.~8, no.~4, pp. 691--717, December 2019.

\bibitem{Unnikrishnan2018}
J.~{Unnikrishnan}, S.~{Haghighatshoar}, and M.~{Vetterli}, ``Unlabeled sensing
  with random linear measurements,'' \emph{IEEE Transactions on Information
  Theory}, vol.~64, no.~5, pp. 3237--3253, May 2018.

\bibitem{Haghighatshoar2018_2}
S.~{Haghighatshoar} and G.~{Caire}, ``Signal recovery from unlabeled samples,''
  \emph{IEEE Transactions on Signal Processing}, vol.~66, no.~5, pp.
  1242--1257, March 2018.

\bibitem{Zhang2019}
H.~{Zhang}, M.~{Slawski}, and P.~{Li}, ``Permutation recovery from multiple
  measurement vectors in unlabeled sensing,'' in \emph{Proceedings of the 2019
  IEEE International Symposium on Information Theory (ISIT)}, July 2019, pp.
  1857--1861.

\bibitem{dokmanic2019permutations}
I.~Dokmani{\'c}, ``Permutations unlabeled beyond sampling unknown,'' \emph{IEEE
  Signal Processing Letters}, vol.~26, no.~6, pp. 823--827, April 2019.

\bibitem{tsakiris19a}
M.~Tsakiris and L.~Peng, ``Homomorphic sensing,'' in \emph{Proceedings of the
  36th International Conference on Machine Learning (ICML)}, vol.~97, June
  2019, pp. 6335--6344.

\bibitem{tsakiris2018eigenspace}
M.~C. Tsakiris, ``Eigenspace conditions for homomorphic sensing,''
  \emph{arXiv:1812.07966}, April 2019.

\bibitem{Dytso2019}
A.~{Dytso}, M.~{Cardone}, M.~S. {Veedu}, and H.~{V. Poor}, ``On estimation
  under noisy order statistics,'' in \emph{Proceedings of the 2019 IEEE
  International Symposium on Information Theory (ISIT)}, July 2019, pp. 36--40.

\bibitem{Kay1998}
S.~M. Kay, \emph{Fundamentals of Statistical Signal Processing, vol. 2:
  Detection Theory}.\hskip 1em plus 0.5em minus 0.4em\relax Prentice Hall PTR,
  1998.

\bibitem{spacingPyke}
\BIBentryALTinterwordspacing
R.~Pyke, ``Spacings,'' \emph{Journal of the Royal Statistical Society. Series B
  (Methodological)}, vol.~27, no.~3, pp. 395--449, 1965. [Online]. Available:
  \url{http://www.jstor.org/stable/2345793}
\BIBentrySTDinterwordspacing

\bibitem{PIT}
J.~E. Angus, ``The probability integral transform and related results,''
  \emph{SIAM review}, vol.~36, no.~4, pp. 652--654, 1994.

\bibitem{boucheron2013concentration}
S.~Boucheron, G.~Lugosi, and P.~Massart, \emph{Concentration inequalities: {A}
  nonasymptotic theory of independence}.\hskip 1em plus 0.5em minus 0.4em\relax
  Oxford university press, 2013.

\bibitem{orderstat}
H.~A. David and H.~N. Nagaraja, ``Order statistics,'' \emph{Encyclopedia of
  statistical sciences}, 2004.

\bibitem{joag1983negative}
K.~Joag-Dev and F.~Proschan, ``Negative association of random variables with
  applications,'' \emph{The Annals of Statistics}, pp. 286--295, 1983.

\end{thebibliography}

\end{document}